\documentclass{JHEP3}

\usepackage{amsfonts}
\usepackage{amsmath}
\usepackage{amssymb}
\usepackage{amsmath,amssymb,epsfig}
\numberwithin{equation}{section}
\usepackage{cite}

\newcommand{\be}{\begin{equation}}
\newcommand{\bea}{\begin{eqnarray}}
\newcommand{\eea}{\end{eqnarray}}
\newcommand{\ba}{\begin{array}}
\newcommand{\ea}{\end{array}}
\newcommand{\ee}{\end{equation}}

\title{Hidden Conformal Symmetry of the Kerr-Newman Black Hole}

\author{Yong-Qiang Wang and Yu-Xiao Liu \\
Institute of Theoretical Physics, Lanzhou
University\\
 Lanzhou 730000, P. R. China\\
 E-mail: \email{yqwang@lzu.edu.cn}, \email{liuyx@lzu.edu.cn}}

\abstract{We investigate the hidden conformal symmetry  of the
4-dimensional non-extremal Kerr-Newman (KN) black hole with
the idea of the near-region Kerr/CFT correspondence proposed
by Castro, Maloney and Strominger in
arXiv:1004.0996[hep-th]. The near-region KN black hole is
dual to a 2D CFT with left and right temperatures $T_L=(2
M^2-Q^2)/(4\pi J)$ and $T_R=\sqrt{M^4-J^2-M^2Q^2}/(2\pi
J)$. Furthermore, we reproduce the microscopic entropy of
the KN black hole via the Cardy formula, which
is in agreement with the macroscopic
Bekenstein-Hawking entropy  and
precisely match the absorption cross section of a
near-region scalar field in the KN black hole with the
finite-temperature absorption cross section for a 2D CFT.}

\begin{document}

\section{Introduction}\label{Sec1}

The study of rotating extremal black holes has been an
interesting subject since the original work on a new
conjecture called the Kerr/CFT correspondence
\cite{KerrCFT0809.4266}, which states that near-horizon
states of a four-dimensional extremal Kerr black hole
could be identified with a two-dimensional chiral CFT on
the spatially infinite boundary, and the central charge is
proportional to the angular momentum of the black hole.
Thus, one can derive the microscopic entropy of the
four-dimensional extremal Kerr black hole by the  Cardy formula.
It is a remarkable fact that this correspondence  does not
rely on supersymmetry or string theory. It was initiated in
the context of the 4D Kerr in general relativity, then
extended to the higher-dimensional rotating solutions in
supergravity and string theories \cite{strings}. More works
on the Kerr/CFT correspondence and its generalization are
listed in \cite{KerrCFT}.

Recently, Castro, Maloney and Strominger (CMS)
\cite{CMS2010a} presented a remarkable result that hidden
conformal symmetry of the 4D non-extremal Kerr Black hole
can be obtained by the study on the low-frequency wave
equation of a massless scalar field in the near region of
the Kerr black hole. A CFT with certain central charges and
temperatures is dual to the corresponding  Kerr Black hole.
This evidence can be verified  by interpreting the
absorption cross section of the scalar in the near region as
finite-temperature absorption section for a 2D CFT.
Subsequently, with the idea of hidden conformal structure
of the non-extremal black hole, Krishnan \cite{Krishnan}
has extended this method to include five-dimensional black
holes in string theory. Furthermore, the non-extremal
uplifted 5D Reissner-Nordstrom black holes have been
investigated by Chen and Sun \cite{ChenSun}.

In 4D general relativity theory, the Kerr black hole can be
considered as the uncharged Kerr-Newman (KN) black hole.
So, we can mention that does the hidden conformal
symmetry of the KN black hole exit, and if so, how? It is
the aim of this article to further investigate the hidden
conformal symmetry of the four-dimensional KN black hole
in general relativity theory following the approach of CMS
in \cite{CMS2010a}. We will study the wave equation of the
near-region scalar field and the corresponding $SL(2,R)$
Casimir structure, and look for the agreement for
absorption cross sections between the CFT and gravity.

The paper is organized as follows. In Sec. \ref{Sec2}, we
review the KN black hole and derive the near-region wave
equation of a massless scalar in the KN background.
Furthermore, with the definition of conformal coordinates,
we obtain the $SL(2,R)$ Casimir structure of wave equation
and reproduce the corresponding microscopic entropy of the KN black hole via the Cardy formula.
In Sec. \ref{secScalarAbsorption}, we show that the
absorption cross section for the near-region scalar field is dual
to the finite temperature absorption cross section in a 2D
CFT. The last section is devoted to conclusion.

\section{Wave equation in the Kerr-Newman background}
 \label{Sec2}

In this section, we study the symmetry of the near-region scalar wave equation in the KN black hole background. The KN solution
describes both the stationary axisymmetric asymptotically flat gravitational
field outside a massive rotating body and a rotating black hole with mass, charge and
angular momentum.

In terms of Boyer--Lindquist coordinates, the KN metric reads
\begin{eqnarray}
 ds^{2} &=& - \left(\frac{\Delta - a^2 \sin^2\theta}{\varrho}\right) dt^2
             - \left(\frac{2 a \sin^2 \theta (r^2+ a^2 -\Delta)}{\varrho}\right) dt d \phi\nonumber\\
   &&+ \left(\frac{(r^2+a^2)^2 - \Delta a^2 \sin^2
        \theta}{\varrho}\right) \sin^2 \theta d \phi^2 + \frac{\varrho}{\Delta}\ dr^2
        + \Sigma\ d\, \theta^2  ,
\end{eqnarray}
where $\Delta$ is the KN horizon function,
\begin{eqnarray}
 \Delta &=& r^{2}-2Mr +a^{2}+Q^2, \nonumber\\
 \varrho &=& r^{2} + a^{2}\cos^{2}\theta,
\end{eqnarray}
and the gauge field $A$ is
\begin{equation}
A=-\frac{Q r}{\varrho}(dt-\sin^2\theta d\phi).
\end{equation}
Here $a$ is the angular momentum for unit mass as measured
from the infinity, and $M$ and $Q$ are the mass and electric
charge of the KN black hole, respectively. The non-extremal
KN black hole has the event horizon $r_{+}$ and the Cauchy
horizon $r_{-}$ at
\begin{equation}
r_{\pm}=M \pm \sqrt{M^2-a^2-Q^2}.
\end{equation}
In what follows, we will see it is convenient to write the
entropy $S_{BH}$, surface gravities $\kappa_{\pm}$, Hawking
temperature $T_H$, angular velocity $\Omega_H$ as :
\begin{eqnarray}
S_{BH } &=&\frac{{\cal A_+}}{4}=  \pi (r_+^2 + a^2), \nonumber\\
\kappa_\pm &=& \frac{2\pi (r_+ - r_-)}{{\cal A_\pm}},\nonumber\\
T_H  &=&\frac{\kappa_+}{2\pi} ={r_+ - r_-\over 4\pi(r_+^2+ a^2)}, \nonumber\\
\Omega_H &=&\frac{4\pi a}{{\cal A_+}}= \frac{a}{r_+^2 + a^2} ,
\end{eqnarray}
here, we use the subscripts ``$\pm$'' to denote the outer and inner horizons of the KN
black hole.

Now, we consider the Klein-Gordon equation for a massless uncharged scalar field in the background of the KN
black hole:
 \begin{eqnarray}
 {\Box}\Phi=0. \label{KGeq}
 \end{eqnarray}
The wave function is written in the eigenmodes of the asymptotic energy $\omega$ and
angular momenta $m$ as
\begin{equation}
  \Phi(t,r,\theta,\phi)=e^{-iwt+ im\phi}R(r)S^\ell(\theta).
\end{equation}
Then, the wave equation (\ref{KGeq}) separates into the angular part
 \begin{equation}\label{angular}
  \left[{1\over \sin\theta}\partial_\theta(\sin\theta\partial_\theta)
   - {m^2\over \sin^2\theta}  + a^2\omega^2 \cos^2\theta \right]
    S^\ell(\theta) = -K_{\ell}S^\ell(\theta),
\end{equation}
and the radial part
\begin{equation}
\left[\partial_r\Delta \partial_r + F + G  \right] R(r) =
K_{\ell} R(r),
  \label{RadialEq}
\end{equation}
with
\begin{eqnarray}
 F &=& \frac{2 a m (Q^2 - 2 M r) \omega  + a^2 (m^2 - 4 M^2 \omega^2)}
       {r (r-2M)+a^2 + Q^2}
     \nonumber \\
   && +\frac{ (Q^4-4 M^2 Q^2 + 8 M^3 r -4 M Q^2 r) \omega^2 }
       {r (r-2M)+a^2 + Q^2} ,
\end{eqnarray}
and
\begin{eqnarray}
 G=(r^2 + 2 M r+4 M^2 - Q^2) \omega^2.
\end{eqnarray}
For $a \omega \ll 1 $, the angular equation (\ref{angular})
can be  written as the spherical harmonic function equation
and the  eigenvalues $K_\ell$ is determined by
\begin{equation}
 K_\ell= \ell(\ell+1).
\end{equation}

Following the argument of CMS in \cite{CMS2010a}, we also
consider the same near-region, which is defined by
\begin{equation}
  \omega M \ll 1 ~~\text{and}~~  r \omega \ll 1. \label{near-region1}
\end{equation}
Then using the condition $M^2 \geq a^2+ Q^2$ and the above equation, we can obtain
\begin{equation}
    \omega Q \ll 1, ~~~Q/r \ll 1. \label{near-region2}
\end{equation}
With the conditions of (\ref{near-region1}) and
(\ref{near-region2}), the term $G$ in Eq.~(\ref{RadialEq})
can be omitted. Thus the radial part of the wave equation
can be reduced to the following near region equation:
\begin{equation}
  \left(\partial_r\Delta \partial_r + F   \right) R(r) = K_{\ell}
  R(r). \label{NearRegionEq}
\end{equation}

Next, we show that Eq.~(\ref{NearRegionEq}) can be
reproduced by the use of the $SL(2,R)$ Casimir operator. First, we
introduce the conformal coordinates ($w^\pm$, $y$),
which is analogue of the definitions in
\cite{CMS2010a,Krishnan,ChenSun}:
\begin{eqnarray}
 w^+ &=&\sqrt{\frac{r-r_+}{ r-r_-}}\, e^{2\pi T_R \phi-2 \lambda_R t},\nonumber\\
 w^- &=&\sqrt{\frac{r-r_+}{ r-r_-}}\,e^{2\pi T_L \phi- 2 \lambda_L t},\nonumber\\
 y &=&\sqrt{\frac{r_+-r_-}{r-r_-}}\, e^{\pi (T_L+T_R) \phi
       - (\lambda_R+\lambda_L)t },
\end{eqnarray}
where
\begin{eqnarray}
  T_R&=&\frac{\kappa_+}{2\pi\Omega_H},~~~~
  T_L=\frac{\kappa_+(\kappa_-+\kappa_+)}
      {2\pi\Omega_H (\kappa_--\kappa_+)},  \label{TRTL} \nonumber\\
  \lambda_R&=&0,~~~~~~~~~~~
  \lambda_L=\frac{\kappa_+ \kappa_-}{\kappa_--\kappa_+}.
\end{eqnarray}
Following the definition of the local vector fields in
\cite{CMS2010a,Krishnan,ChenSun}:
\begin{eqnarray}
 H_1 &=& i \partial_+,~~\nonumber\\
 H_0 &=& i \big( w^+ \, \partial_+ + \frac12 y \, \partial_y \big), \\
 H_{-1} &=& i \left( w^{+2} \, \partial_+ + w^+ y \, \partial_y - y^2 \, \partial_- \right),\nonumber
\end{eqnarray}
and
\begin{eqnarray}
 \bar H_1 &=& i \partial_-,~~ \nonumber\\
 \bar H_0 &=&i \big( w^- \, \partial_- + \frac12 y \, \partial_y \big), \\
\bar H_{-1} &=& i \left( w^{-2} \, \partial_- + w^- y \, \partial_y - y^2 \,\nonumber
\partial_+ \right),
\end{eqnarray}
we can obtain the two sets of $SL(2,R)$ Lie algebra with respect to $(H_0,{ H}_{\pm
1})$ and $({\bar H},{\bar H}_{\pm 1})$, respectively:
\begin{equation}
 \left[ H_0, H_{\pm 1} \right] = \mp i H_{\pm 1}, \qquad
 \left[ H_{-1}, H_1 \right] = - 2 i H_0,
\end{equation}
\begin{equation}
\left[{\bar H}_0, {\bar H}_{\pm 1} \right] = \mp i {\bar H}_{\pm 1}, \qquad \left[{\bar H}_{-1}, {\bar H}_1 \right] = - 2 i {\bar H}_0.
\end{equation}
The corresponding  quadratic Casimir operator of the
$SL(2,R)$ Lie algebra reads as
\begin{eqnarray}
 \mathcal{H}^2 &=& \mathcal{\bar H}^2
     = - H_0^2 + \frac{1}{2} \left( H_1 H_{-1}
       + H_{-1} H_1 \right) \nonumber\\
  &=& \frac{1}{4} \left(y^2 \, \partial_y^2 - y \,\partial_y \right)
      + y^2 \, \partial_+\partial_-.
\end{eqnarray}
Thus, we rewrite the vector fields in terms of the coordinators $(t,r ,\phi)$ as
\begin{eqnarray}
  H_1 &=&i e^{-2\pi  T_R \phi }
      \left(\sqrt{\Delta}\partial _r
            +\frac{1}{2{\pi}T_R}\frac{r-\frac{1}{4\lambda_L} }
                 {\sqrt{\Delta}}\partial _{\phi }
            +\frac{2 T_L}{ T_R}\frac{M r -\frac{(a^2 + M^2) Q^2}{2 M^2 - Q^2}-a^2}
                               {\sqrt{\Delta}}\partial_t\right),  \nonumber\\
   H_0 &=&\frac{i}{2\pi T_R}\partial_\phi
          +2 i \frac{1}{4\lambda_L}\frac{T_L}{T_R}\partial_t, \\
  H_{-1}&=&i e^{2\pi  T_R \phi }
       \left(-\sqrt{\Delta}\partial _r
       +\frac{1}{2\pi  T_R}\frac{r-\frac{1}{4\lambda_L} }
              {\sqrt{\Delta}}\partial _{\phi }
              +\frac{2 T_L}{ T_R}\frac{M r-\frac{(a^2 + M^2) Q^2}{2 M^2 - Q^2}-a^2}
              {\sqrt{\Delta}}\partial _t\right),\nonumber
\end{eqnarray}
and
\begin{eqnarray}
\bar H_1 &=&i e^{-2\pi  T_L \phi
     +2 \lambda_L t}\left(\sqrt{\Delta}\partial_r
     -\frac{a}{\sqrt{\Delta}}\partial _{\phi}
    -(2M-\frac{Q^2}{r})\frac{r}
    {\sqrt{\Delta}}\partial _t\right), \nonumber\\
 \bar H_0 &=&-2 i M \partial_t, \\
 \bar H_{-1}&=&i e^{2\pi  T_L \phi
     -2 \lambda_L t}\left(- \sqrt{\Delta}\partial _r
     -\frac{a}{\sqrt{\Delta}}\partial _{\phi }
     -(2 M-\frac{Q^2}{r})\frac{r}{ \sqrt{\Delta}}\partial _t\right).\nonumber
\end{eqnarray}
So, we obtain the corresponding Casimir operator
\begin{eqnarray}\mathcal{H}^2&=&
    \partial_r\Delta \partial_r
    +\frac{2 a m (Q^2 - 2 M r) \omega  + a^2 (m^2 - 4 M^2 \omega^2)}
       {r (r-2M)+a^2 + Q^2}
     \nonumber \\
   && +\frac{ (Q^4-4 M^2 Q^2 + 8 M^3 r -4 M Q^2 r) \omega^2 }
       {r (r-2M)+a^2 + Q^2} .
\end{eqnarray}
In the background of a KN black hole, the near region wave
equation (\ref{NearRegionEq}) can be rewritten as
\begin{equation}
   \mathcal{\bar H}^2 \Phi = \mathcal{H}^2 \Phi = l (l+1) \Phi.
\end{equation}

From the above procedure, we know that, in the near region,
the scalar field can have $SL(2,R)_L \times SL(2,R)_R $
weight to obtain the microscopical entropy of the
non-extremal KN black hole by the use of the Cardy Formula. However,
we don't know how to compute the corresponding central
charges $c_L$ and $c_R$. Here, as did in \cite{CMS2010a}, we
also assume that the central charges of the non-extremal
Kerr-Newman still keep same as the extremal case, i.e.,
$c_L=c_R=12J$. Then, using the Cardy formula for the
microstate, we have
 \begin{equation}
S = \frac{\pi^2}3 (c_L T_L + c_R T_R) = \pi ( r_+^2 + a^2) .
\end{equation}
From the 2D dual CFT, we reproduce the KN black hole
entropy in four dimensions which reaches agreement with the
macroscopic entropy $S_{BH}$.

\section{Scalar absorption}\label{secScalarAbsorption}

In this section we will calculate the  absorption cross
section for the KN black hole in the near region. The
absorption cross section of a low-frequency massless scalar
in the near-extremal limit of the KN black hole has been
calculated and analyzed in
\cite{MaldacenaIH,CveticXV,CveticAP}. In the near region,
the solutions of the radial wave equation
(\ref{NearRegionEq}) with respect to the ingoing and outgoing
boundary conditions at the horizon  are
\begin{eqnarray}
 R_{in}(r)&=&\left(\frac{r-r_+}{ r-r_-}\right)
             ^{-i\frac{(\omega-m\Omega_H)}{4\pi T_H}}
             (r-r_-)^{-1-\ell}\nonumber\\
 &&{\times}F\left(1+\ell-i\frac{4M^2-2Q^2}{ r_+-r_-}\omega
               +i\frac{m\Omega_H}{ 2\pi T_H},
          1+\ell-i2M{\omega};
          1-i\frac{(\omega-m\Omega_H)}{ 2\pi T_H};
          \frac{r-r_-}{ r-r_+}\right), \nonumber\\
\end{eqnarray}
and
\begin{eqnarray}
 R_{out}(r)&=&\left(\frac{r-r_+}{ r-r_-}\right)
              ^{i\frac{(\omega-m\Omega_H)}{4\pi T_H}}
              (r-r_-)^{-1-\ell} \nonumber \\
 &&{\times}F\left(1+\ell+i\frac{4M^2-2Q^2}{ r_+-r_-}\omega
             -i\frac{m\Omega_H}{ 2\pi T_H},
          1+\ell+i2M{\omega};
          1+i\frac{(\omega-m\Omega_H)}{ 2\pi T_H};
          \frac{r-r_-}{ r-r_+}\right),\nonumber \\
\end{eqnarray}
where $F(a,b; c;z)$ is the hypergeometric function. At the
outer boundary of the matching region  $r\gg M$ (but still
$r\ll {1 \over\omega}$) , the ingoing wave behaves as
\begin{equation}
  R_{in}(r\gg M)\sim Ar^\ell
\end{equation}
with
\begin{equation}
  A={\Gamma(1-i\frac{\omega-m\Omega_H}{ 2\pi T_H})
    \Gamma(1+2\ell)\over\Gamma(1+\ell-i{2M\omega})\,
    \Gamma(1+\ell-i\frac{4M^2-2Q^2}{ r_+-r_-}\omega
           +\frac{i}{2\pi T_H}m\Omega_H)}~.
\end{equation}
The absorption cross section can be  written as
\begin{eqnarray}
 P_{\rm abs}&&\sim |A|^{-2}\nonumber \\
 && \sim \sinh\left(\frac{\omega-m\Omega}{2 T_H}\right)
    \left|\Gamma\left(1+\ell-i{2M\omega}\right)\right|^2\nonumber\\
 && \quad\,\times\left|\Gamma
   \left(1+\ell-i\frac{4M^2-2Q^2}{ r_+-r_-}\omega
         +\frac{i}{ 2\pi T_H}m\Omega_H\right)\right|^2.
\end{eqnarray}
In order to match the absorption cross section of a
near-region scalar field in the KN black hole background
with the finite-temperature absorption cross section for
the corresponding 2D CFT, we need study the first law of
black hole thermodynamics. Since the existence of non-zero
charge of the KN black hole, the first law of black hole
thermodynamics is
\begin{equation}
    T_{H}\delta S=\delta M -\Omega_H \delta J-\varphi \delta Q,
\end{equation}
where $\varphi= \frac{Q ~r_+}{r_+^2+ a^2} $ is the electric
potential of the KN black hole. It is obvious that $\delta
Q$ is equal to  zero  with respect to the uncharged scalar field. With the conjugate charges $\delta E_R$ and
$\delta E_L$, we can suppose the dual CFT entropy is
\begin{equation}
    \delta S={\delta E_L \over T_L}+{\delta E_R \over T_R}.
\end{equation}
The solution is
\begin{eqnarray}
 \delta E_L&=& \frac{2M^2-Q^2}{ a}\delta M~,\nonumber\\
 \delta E_R &=& \frac{2M^2-Q^2}{ a}\delta M-\delta J~.
\end{eqnarray}
With the variations $\delta M =m$ and $\delta J =\omega$,
the left and right moving frequencies are turned out to be
\begin{eqnarray}
 \omega_L &\equiv& \delta E_L= \frac{2M^2-Q^2}{a}\omega~,\nonumber\\
 \omega_R &\equiv&  \delta E_R =\frac{2M^2-Q^2}{a}\omega-m.
\end{eqnarray}
By using of the above formula, we can rewrite the
absorption cross section as
\begin{equation}
  P_{\rm abs}\sim T_L^{2h_L-1}T_R^{2h_R-1}
      \sinh\left({\omega_L\over 2T_L}+{\omega_R \over 2 T_R}\right)
      \left|\Gamma(h_L+i{\omega_L \over 2\pi T_L})\right|^2
      \left|\Gamma(h_R+i{\omega_R\over 2\pi T_R})\right|^2~,
\end{equation}
which agrees precisely with finite-temperature absorption
cross section for a 2D CFT.

\section{Conclusion}\label{SecConclusions}

In this paper, the method of CMS is extended to the four-dimensional KN Black Hole in general relativity theory. We
investigate the hidden conformal structure of the KN black
hole by the low-frequency wave equation of a massless
scalar in the near region. The dual CFT with the left and right
temperatures $T_L=(2
M^2-Q^2)/(4\pi J)$ and $T_R=\sqrt{M^4-J^2-M^2Q^2}/(2\pi
J)$ is obtained.
Furthermore, under the assumption that the central charges
of the non-extremal Kerr-Newman still keep same as the
extremal case, one can reproduce the macroscopic
Bekenstein-Hawking entropy via the Cardy formula. At last, by rewriting the
absorption cross section of a near-region scalar field as
the finite-temperature absorption cross section for the 2D
CFT, one can verify that the
near-region KN black hole is dual to a 2D CFT.

From the result of this paper, we can see that the method
of CMS is valid for the non-extremal Kerr-Newman black hole in 4D
general relativity theory background. It would be interesting to
investigate other charged and rotating black holes in
diverse dimensions and in various gravity theories.

\section*{Acknowledgment}

We would like to thank Zhen-Hua Zhao for discussions. This
work was partly supported by the Program for New Century
Excellent Talents in University, the National Natural
Science Foundation of China (No. 10705013), the Doctoral
Program Foundation of Institutions of Higher Education of
China (No. 20090211110028), the Key Project of Chinese
Ministry of Education (No. 109153), and the Fundamental
Research Funds for the Central Universities (No.
lzujbky-2009-54 and No. lzujbky-2009-122).


%


\begin{thebibliography}{99}
\bibitem{KerrCFT0809.4266}
 M.~Guica, T.~Hartman, W.~Song and A.~Strominger,
    ``The Kerr/CFT Correspondence,''
    Phys. Rev. D \textbf{80}, 124008 (2009)
    [arXiv:0809.4266 [hep-th]].


\bibitem{strings}
M.~Cvetic and F.~Larsen,
    ``General rotating black holes in string theory:
      Greybody factors and event horizons,''
    Phys.\ Rev.\  D {\bf 56}, 4994 (1997)
    [arXiv:hep-th/9705192];
M.~Cvetic and D.~Youm,
    ``General Rotating Five Dimensional Black Holes
      of Toroidally Compactified Heterotic String,''
    Nucl.\ Phys.\  B {\bf 476}, 118 (1996)
    [arXiv:hep-th/9603100];
S.~W.~Hawking, C.~J.~Hunter and M.~Taylor,
    ``Rotation and the AdS/CFT correspondence,''
    Phys.\ Rev.\  D {\bf 59}, 064005 (1999)
    [arXiv:hep-th/9811056];
H.~Lu, J.~Mei, C.~N.~Pope and J.~Vazquez-Poritz,
    ``Extremal Static AdS Black Hole/CFT Correspondence
      in Gauged Supergravities,''
    Phys.~Lett.~B \textbf{673}, 77 (2009)
    [arXiv:0901.1677 [hep-th]];
D.~D.~K.~Chow, M.~Cvetic, H.~Lu and C.~N.~Pope,
    ``Extremal Black Hole/CFT Correspondence in (Gauged) Supergravities,''
    Phys.~Rev.~D \textbf{79}, 084018 (2009)
    [arXiv:0812.2918 [hep-th]];
A.~M.~Ghezelbash,
    ``Kerr/CFT Correspondence in Low Energy Limit
      of Heterotic String Theory,''
    JHEP \textbf{0908}, 045 (2009)
    [arXiv:0901.1670 [hep-th]];
T.~Azeyanagi, N.~Ogawa and S.~Terashima,
    ``The Kerr/CFT Correspondence and String Theory,''
    Phys.~Rev.~D \textbf{79}, 106009 (2009)
    [arXiv:0812.4883 [hep-th]];
W.~Y.~Wen,
    ``Holographic descriptions of (near-)extremal black holes
      in five dimensional minimal supergravity,''
    arXiv:0903.4030 [hep-th];
C.~M.~Chen and J.~E.~Wang,
    ``Holographic Duals of Black Holes in
    Five-dimensional Minimal Supergravity,''
    arXiv:0901.0538 [hep-th].


\bibitem{KerrCFT}
D. Astefanesei and H. Yavartanoo,
    ``Stationary black holes and attractor mechanism,''
    Nucl.~Phys.~B \textbf{794}, 13 (2008)
    [arXiv:0706.1847 [hep-th]];
R.~A.~Konoplya and A.~Zhidenko,
    ``Decay of a charged scalar and Dirac fields in
      the Kerr-Newman-de Sitter background,''
    Phys.~Rev.~D {\bf 76}, 084018 (2007)
    [arXiv:0707.1890 [hep-th]];
B.~Chen and Z.~b.~Xu,
    ``Quasi-normal modes of warped black holes
      and warped AdS/CFT correspondence,''
    JHEP {\bf 0911}, 091 (2009)
    [arXiv:0908.0057 [hep-th]];
B.~Chen, B.~Ning and Z.~b.~Xu,
    ``Real-time correlators in warped AdS/CFT correspondence,''
    JHEP {\bf 1002}, 031 (2010)
    [arXiv:0911.0167 [hep-th]];
B.~Chen and C.~S.~Chu,
    ``Real-time correlators in Kerr/CFT correspondence,''
    arXiv:1001.3208 [hep-th];
T.~Hartman, K.~Murata, T.~Nishioka and A.~Strominger,
    ``CFT Duals for Extreme Black Holes,''
    JHEP \textbf{0904}, 019 (2009)
    [arXiv:0811.4393 [hep-th]];
D. Astefanesei and Y. K. Srivastava,
    ``CFT Duals for Attractor Horizons,''
    Nucl.~Phys.~B \textbf{822}, 283 (2009)
    [arXiv:0902.4033 [hep-th]];
A.~M.~Ghezelbash,
    ``Kerr-Bolt Spacetimes and Kerr/CFT Correspondence,''
    arXiv:0902.4662 [hep-th];
M.~R.~Garousi and A.~Ghodsi,
    ``The RN/CFT Correspondence,''
    Phys.~Lett.~B \textbf{687}, 79 (2010)
    [arXiv:0902.4387 [hep-th]];
K.~Hotta,
    ``Holographic RG flow dual to attractor flow
      in extremal black holes,''
    Phys.~Rev.~D \textbf{79}, 104018 (2009)
    [arXiv:0902.3529 [hep-th]];
T.~Fukuyama,
    ``SO(2,d-1) Gauge Theory of Gravity in d Dimensional
      Spacetime and $AdS_d/CFT_{d-1}$ Correspondence,''
    arXiv:0902.2820 [hep-th];
S.~Nam and J.~D.~Park,
    ``Hawking radiation from covariant anomalies
      in 2+1 dimensional black holes,''
    Class.~Quant.~Grav.~\textbf{26}, 145015 (2009)
    [arXiv:0902.0982 [hep-th]];
B.~Chen and Z.~b.~Xu,
    ``Quasinormal modes of warped $AdS_3$ black holes
      and AdS/CFT correspondence,''
    arXiv:0901.3588 [hep-th];
F.~Loran and H.~Soltanpanahi,
    ``5D Extremal Rotating Black Holes and CFT duals,''
    Class.~Quant.~Grav.~\textbf{26}, 155019 (2009)
    [arXiv:0901.1595 [hep-th]];
J.~J.~Peng and S.~Q.~Wu,
    ``Extremal Kerr black hole/CFT correspondence
      in the five dimensional G\'odel universe,''
    Phys.~Lett.~B \textbf{673}, 216 (2009)
    [arXiv:0901.0311 [hep-th]];
H.~Isono, T.~S.~Tai and W.~Y.~Wen,
    ``Kerr/CFT correspondence and five-dimensional BMPV black holes,''
    Int.~J.~Mod.~Phys.~A \textbf{24}, 5659 (2009)
    [arXiv:0812.4440 [hep-th]];
Y.~Nakayama,
    ``Emerging AdS from Extremally Rotating NS5-branes,''
    Phys.~Lett.~B \textbf{673}, 272 (2009)
    [arXiv:0812.2234 [hep-th]];
A.~Garbarz, G.~Giribet and Y.~Vasquez,
    ``Asymptotically AdS3 Solutions to Topologically
      Massive Gravity at Special Values of the Coupling Constants,''
    Phys.~Rev.~D \textbf{79}, 044036 (2009)
    [arXiv:0811.4464 [hep-th]];
T.~Azeyanagi, N.~Ogawa and S.~Terashima,
    ``Holographic Duals of Kaluza-Klein Black Holes,''
    JHEP \textbf{0904}, 061 (2009)
    [arXiv:0811.4177 [hep-th]];
 G.~W.~Gibbons, C.~A.~R.~Herdeiro, C.~M.~Warnick and M.~C.~Werner,
    ``Stationary Metrics and Optical Zermelo-Randers-Finsler Geometry,''
    Phys.~Rev.~D \textbf{79}, 044022 (2009)
    [arXiv:0811.2877 [gr-qc]];
 K.~Hotta, Y.~Hyakutake, T.~Kubota, T.~Nishinaka and H.~Tanida,
    ``The CFT-interpolating Black Hole in Three Dimensions,''
    JHEP {\bf 0901}, 010 (2009)
    [arXiv:0811.0910 [hep-th]];
 M.~Schvellinger,
    ``Kerr-AdS black holes and non-relativistic conformal
      QM theories in diverse dimensions,''
    JHEP {\bf 0812}, 004 (2008)
    [arXiv:0810.3011 [hep-th]];
 M.~Becker, S.~Cremonini and W.~Schulgin,
    ``Extremal Three-point Correlators in Kerr/CFT,''
    arXiv:1004.1174 [hep-th];
 J.~Mei,
    ``The Entropy for General Extremal Black Holes,''
    JHEP {\bf 1004}, 005 (2010)
    [arXiv:1002.1349 [hep-th]];
 C.~M.~Chen, Y.~M.~Huang and S.~J.~Zou,
    ``Holographic Duals of Near-extremal Reissner-Nordstrom Black Holes,''
    JHEP {\bf 1003}, 123 (2010)
    [arXiv:1001.2833 [hep-th]];
 J.~J.~Peng and S.~Q.~Wu,
    ``Extremal Kerr/CFT correspondence of five-dimensional
      rotating (charged) black holes with squashed horizons,''
    Nucl.~Phys.~B {\bf 828}, 273 (2010)
    [arXiv:0911.5070 [hep-th]];
 C.~Krishnan,
    ``Tomograms of Spinning Black Holes,''
    Phys.~Rev.~D {\bf 80}, 126014 (2009)
    [arXiv:0911.0597 [hep-th]];
 C.~M.~Chen, J.~R.~Sun and S.~J.~Zou,
    ``The RN/CFT Correspondence Revisited,''
    JHEP {\bf 1001}, 057 (2010)
    [arXiv:0910.2076 [hep-th]];
 J.~Rasmussen,
    ``A note on Kerr/CFT and free fields,''
    arXiv:0909.2924 [hep-th];
 D.~Grumiller and A.~M.~Piso,
    ``Exact relativistic viscous fluid solutions
      in near horizon extremal Kerr background,''
    arXiv:0909.2041 [astro-ph.SR];
 J.~Rasmussen,
    ``Isometry-preserving boundary conditions
      in the Kerr/CFT correspondence,''
    Int.~J.~Mod.~ Phys.~A {\bf 25}, 1597 (2010)
    [arXiv:0908.0184 [hep-th]];
 Y.~Matsuo, T.~Tsukioka and C.~M.~Yoo,
    ``Yet Another Realization of Kerr/CFT Correspondence,''
    Europhys.~Lett.~{\bf 89}, 60001 (2010)
    [arXiv:0907.4272 [hep-th]];
 Y.~Matsuo, T.~Tsukioka and C.~M.~Yoo,
    ``Another Realization of Kerr/CFT Correspondence,''
    Nucl.~Phys.~B {\bf 825}, 231 (2010)
    [arXiv:0907.0303 [hep-th]];
 O.~J.~C.~Dias, H.~S.~Reall and J.~E.~Santos,
    ``Kerr-CFT and gravitational perturbations,''
    JHEP {\bf 0908}, 101 (2009)
    [arXiv:0906.2380 [hep-th]];
 L.~M.~Cao, Y.~Matsuo, T.~Tsukioka and C.~M.~Yoo,
    ``Conformal Symmetry for Rotating D-branes,''
    Phys.~Lett.~B {\bf 679}, 390 (2009)
    [arXiv:0906.2267 [hep-th]];
 W.~Kim and E.~J.~Son,
    ``Central Charges in 2d Reduced Cosmological Massive Gravity,''
    Phys.\ Lett.\  B {\bf 678}, 107 (2009)
    [arXiv:0904.4538 [hep-th]];
 X.~N.~Wu and Y.~Tian,
    ``Extremal Isolated Horizon/CFT Correspondence,''
    Phys.\ Rev.\  D {\bf 80}, 024014 (2009)
    [arXiv:0904.1554 [hep-th]];
 C.~Krishnan and S.~Kuperstein,
    ``A Comment on Kerr-CFT and Wald Entropy,''
    Phys.\ Lett.\  B {\bf 677}, 326 (2009)
    [arXiv:0903.2169 [hep-th]];
 T.~Azeyanagi, G.~Compere, N.~Ogawa, Y.~Tachikawa and S.~Terashima,
    ``Higher-Derivative Corrections to the Asymptotic
      Virasoro Symmetry of 4d Extremal Black Holes,''
    Prog.\ Theor.\ Phys.\  {\bf 122}, 355 (2009)
    [arXiv:0903.4176 [hep-th]];
M.~Cvetic and F.~Larsen,
    ``Greybody Factors and Charges in Kerr/CFT,''
    JHEP {\bf 0909}, 088 (2009)
    [arXiv:0908.1136 [hep-th]];
T.~Hartman, W.~Song and A.~Strominger,
    ``Holographic Derivation of Kerr-Newman Scattering
      Amplitudes for General Charge and Spin,''
    JHEP \textbf{1003}, 118 (2010)
    [arXiv:0908.3909 [hep-th]];
H.~Lu, J.~Mei and C.~N.~Pope,
    ``Kerr/CFT Correspondence in Diverse Dimensions,''
    JHEP {\bf 0904}, 054 (2009)
    [arXiv:0811.2225 [hep-th]];
I.~Bredberg, T.~Hartman, W.~Song and A.~Strominger,
    ``Black Hole Superradiance From Kerr/CFT,''
    JHEP \textbf{1004}, 019 (2010.)
    [arXiv:0907.3477 [hep-th]].



\bibitem{CMS2010a}
 A.~Castro, A.~Maloney and A.~Strominger,
    ``Hidden Conformal Symmetry of the Kerr Black Hole,''
    arXiv:1004.0996 [hep-th].


\bibitem{Krishnan}
 C.~Krishnan,
    ``Hidden Conformal Symmetries of Five-Dimensional Black Holes,"
    arXiv:1004.3537 [hep-th].


\bibitem{ChenSun}
 C.-M.~Chen and J.-R. Sun,
    ``Hidden Conformal Symmetry of the Reissner-Nordstr{\o}m Black Holes,''
    arXiv:1004.3963 [hep-th].


\bibitem{MaldacenaIH}
  J.~M.~Maldacena and A.~Strominger,
    ``Universal low-energy dynamics for rotating black holes,''
    Phys. Rev. D \textbf{56}, 4975 (1997)
    [arXiv:hep-th/9702015].

\bibitem{CveticXV}
 M.~Cvetic and F.~Larsen,
    ``Greybody factors for rotating black holes in four dimensions,''
    Nucl. Phys. B \textbf{506}, 107 (1997)
    [arXiv:hep-th/9706071].

\bibitem{CveticAP}
 M.~Cvetic and F.~Larsen,
    ``Greybody factors for black holes in four dimensions:
      Particles with spin,''
    Phys. Rev.  D \textbf{57}, 6297 (1998)
    [arXiv:hep-th/9712118].






\end{thebibliography}
\end{document}